# Controlling the macroscopic electrical properties of reduced graphene oxide by nanoscale writing of electronic channels

*Arijit Kayal[1,#], Harikrishnan G[1], K. Bandopadhyay[1]\*, Amit Kumar[2], S. Ravi P. Silva[3], J. Mitra[1,†]*

[1]School of Physics, Indian Institute of Science Education and Research, Thiruvananthapuram 695551, Kerala, India.

[2]School of Mathematics and Physics, Queen's University Belfast, BT7 1NN, United Kingdom.

[3]Advanced Technology Institute, University of Surrey, Guildford, GU2 7XH, United Kingdom.

ABSTRACT

The allure of all carbon electronics stems from the spread in physical properties, across all its allotropes. The scheme also harbours unique challenges, like tunability of band-gap, variability of doping and defect control. Here, we explore the technique of scanning probe tip induced nanoscale reduction of graphene oxide (GO), which nucleates conducting, $sp^2$ rich graphitic regions on the insulating GO background. Flexibility of direct writing is supplemented with control over degree of reduction and tunability of bandgap, through macroscopic control parameters. The fabricated reduced —GO channels and ensuing devices are investigated via spectroscopic, and temperature and bias dependent electrical transport and correlated with spatially resolved electronic properties, using surface potentiometry. Presence of carrier localization effects, induced by the phase-separated $sp^2/sp^3$ domains, and large local electric field fluctuations are reflected in the non-linear transport across the channels. Together the results indicate a complex transport phenomena which may be variously dominated by tunnelling, variable range hopping or activated depending on the electronic state of the material.

INTRODUCTION

Realization of graphene-based electronic devices have been restricted primarily due to the lack of reproducible and cost-effective preparation techniques for graphene and the absence of an electronic bandgap ($E_G$).[1] Superficially, graphene oxide (GO) appears to offer a suitable alternative in terms of scalability following the various modifications of the Hummer's method[2] and also due to the presence of a notional bandgap in its density of states (DOS). Graphene oxide results from addition of oxygen-containing side groups to the hexagonal carbon lattice of graphite via acid functionalization process, and subsequently reducing some of these states with hydrogen, while passivating defects. However, research into controlled defect incorporation into graphene[3-6] has shown that surface adsorbates or doping induced band structure tailoring compromises the ensuing electrical properties and likewise in GO. Like graphene, GO has a layered structure but with significantly larger inter-planar separation due to incorporation of the epoxy groups, which makes it amenable for exfoliation. Incorporation of the oxygenated functional groups severely disrupts the honeycomb lattice of graphene, making GO a very dirty insulator with dubious electrical properties. In reverse, reproducible reduction of GO to graphene has proved to be a non-trivial task. While it has been shown that more than 90% of the oxygen-containing groups can be effectively removed from GO,[7] restoration of long-range $sp^2$ network and the associated electrical and optical properties, akin to those of graphene remains challenging. Since the physical properties of this intermediate material, reduced GO (rGO) lies between those of GO and graphene, any systematic variation of rGOs physical state and properties by controlled reduction is of huge fundamental and technological importance. Though innumerable chemical[8-10] and physical[11-16] reduction methods have been reported, identification of robust and controllable process parameters remain illusive. From a more fundamental viewpoint, the atomic structure of rGO remains unclear, beyond the fact that the $sp^2$ network of graphene is disrupted in GO, creating a disordered lattice where the oxygen functional groups are covalently bonded to the carbon atoms along with some adsorbed oxidative debris. The process of structural rearrangement upon reduction of GO to rGO and then graphene also remains hazy. Thermodynamically, GO would have higher entropy (disorder) compared to graphene (ordered) though energetically GO would be lower than graphene. As a part of the reduction process the oxygen functional groups, which are bonded with the ($sp^3$ hybridized) carbon atoms need be removed from the system allowing the amorphous carbon network to reorganize in the hexagonal lattice. Evolution of the electronic band structure of GO to rGO and then towards graphene, under progressive reduction, also remains unclear.[17] Angle-resolved photoemission spectroscopy of graphene, exposed to atomic oxygen does not display signatures of doping unlike that for atomic nitrogen,[18] but does evidence localization of electronic states in the valence band ($V_B$), especially those close to the Fermi level ($E_F$)[19]. This is similar to the case of atomic *N* bonding to amorphous *C*, which has many non-doping states available via hybridized states.[20] Further, the substrate is known to play a crucial role in determining the observed band structure of rGO, similar to the case of bilayer graphene has been shown to display a dispersionless flat band near the *K* point along with a finite bandgap.[21] The electrical properties of rGO, like carrier density and mobility are also compromised compared to those of graphene with the best rGO samples exhibiting mobility ~ $10^3$ cm$^2$V$^{-1}$s$^{-1}$, along with conductivity comparable to that of polycrystalline graphite.[11] Overall, investigations into electrical and optical properties of rGO show wide variability in properties, allowing multiple inferences to be drawn. Electrical transport in rGO has been variously shown follow thermally activated transport[22] or two-dimensional (2D) variable range hopping (VRH)[23-24]. The later may be dominated either by Mott-VRH or the electric field controlled Efros-Shklovskii (ES-VRH) model.[25-26] Each of the above models are characterized by an exponent ($p$) describing the temperature ($T$) dependence of low bias conductivity ($\sigma$), generically given as $\sigma(T) \propto \exp(T_0/T)^p$, where $T_0$ is a fit parameter characterizing the material. The exponent $p$ assumes the value 1 for activated transport, and 1/3 and ½ for the 2D Mott-VRH and ES-VRH models, respectively. Based on temperature dependent transport measurements alone it is difficult to unequivocally distinguish between the two VRH models, often with data that may reasonably fit either model, in restricted temperature ranges. Such ambiguity is not only restricted to rGO but extends to epitaxially grown monolayer graphene[27] and its amorphous counterpart, hydrogenated amorphous carbon (a-C:H)[28]. Theoretical calculations also show that scattering due to defect induced charge impurities in graphene and the presence of localized states (near $E_F$) in rGO leads to spatial fluctuations in the surface charge density with local accumulation of electrons and holes, limiting the conductivity of the material.[29]

Against this backdrop, we investigate a highly controllable and spatially accurate route to GO "reduction" using a conducting atomic force microscope (C-AFM) based lithography technique that locally reduces GO to rGO.[13-14] Here, repeated scanning of a designated area on monolayer to few-layer GO flakes, by a C-AFM tip under suitable bias and humidity, increases the local electrical conductance of the area, indicating possible local reduction of GO. Scanning probe microscopy (SPM) based lithography techniques have been extensively researched over the last three decades.[30-31] They offer integrated patterning and imaging capabilities, with high spatial resolution, depth of field and operational flexibility i.e. wide range of processable materials[30-31] that are not possible with other techniques. In all cases, the scanning tip induces physical (mechanical, thermal, diffusive etc.) or chemical (oxidation etc.) changes to the surface altering the local properties. The spatial resolution of such changes are dependent on the characteristic length scales of the interaction(s) exploited. In the present context, this is one technique that allows a single-step patterning cum reduction protocol and a possible route towards realizing all carbon electronic devices.[32-33] Importantly, it decreases the use of lithography resists on 2D materials such as graphene, that are known to be sensitive to chemical organic residues. Though the technique has limited throughput, in its present form, it allows a high degree of control on the reduction process through a few highly controllable macroscopic parameters such as tip-sample bias ($V_S$), relative humidity ($RH\%$) and number of area scans or voltage cycles. Further, the throughput limitations will be mitigated by the promise of scan speeds approaching millimetres per second, opening up newer avenues for tip-based material engineering and lithography.

There have been a few reports on C-AFM tip-based local reduction of GO, adopting thermal and electrochemical approaches to local modification of the GO surface.[13-16] While all these studies have been led by the basic aim of customizable and spatially localized nanometer-scale transformation of GO into rGO, the chemical identity and electronic structure of the resulting rGO has largely remained unexplored. In all cases, the primary indicator of "reduction" has been change in local electrical property of the scanned region, with the limited size of the modified regions restricting spectroscopic quantification of their chemical identity. Here, we continue to refer to these modified regions as rGO and investigate their chemical and physical properties, and importantly the dependency of these properties on the macroscopic control parameters, identified earlier. Though the exact mechanism of this tip induced modification is still not fully understood, it has been proposed that desorption of oxygenated functional groups from the GO surface are induced by local electrochemistry, in presence of the tip-sample junction electric field[13-14] and water condensed therein or thermochemistry.[15-16] Apart from a significant increase in the local conductivity of tip induced rGO ($\sim 10^3\ S/m$), which is comparable to those obtained by chemical reduction,[16] little is known about their atomic/electronic structure and hence the carrier transport mechanism, which are investigated here. Available Raman spectroscopy studies over these tip induced rGO domains have shown a relative increase in the fraction of $sp^2$ and $sp^3$ domains and reveal spatial inhomogeneity in the atomic structure.[13-14, 34] Generically, the presence of such spatial inhomogeneity often leads to inhomogeneous electronic properties and as is expected in the case of rGO proposed here. Along with change in topographic and mechanical properties of rGO, we have investigated the effect of reduction through recording the spatial inhomogeneity in local current and local conductance ($dI/dV$) of the surface. rGO devices fabricated by "drawing" a conducting rGO channel between a pair of source-drain electrodes then allowed us to correlate the non-equilibrium, electrical transport properties of the rGO channel with the nanoscale electronic properties of the channel. Evolution of spatially resolved surface potential ($V_S$) maps under zero and non-zero source-drain bias reflects the variation in local electronic structure and quantifies the fluctuations in the local electric field, which are seen to be two orders in magnitude larger than the macroscopic field. Our results support the hypothesis that the observed non-linearity in the device current-voltage ($IV$) characteristics stem from intrinsic material inhomogeneity of the channel and the ensuing charge transfer between localized states, rather than purely from the metal-rGO contacts at either end of the device. The $IV$ nonlinearities are seen to decrease systematically with the degree of reduction of the rGO channel further justifying its origin based on the intrinsic material property that evolves with reduction. The rGO surface inhomogeneities are also visualized by 2D differential voltage ($\nabla^2 V_S$) maps that evidence charge puddles extending over tens of nanometers. Overall, we present a combination of macroscopic and nanoscale electrical investigations characterising the physical properties of rGO devices directly written (reduced) by a C-AFM tip together with micro Raman maps characterising the chemical nature of rGO.

EXPERIMENTAL METHODS

Water dispersion of Graphene oxide (GO) ($2.5\ wt\%$ concentration) sample was purchased from Graphenea. The suspension was further diluted in a $1:20$ volumetric ratio and then spin-coated onto selected substrates for further processing and characterisation. All AFM experiments were performed using the Bruker Multimode-8 AFM in a humidity-controlled environment. Experiments in local reduction of GO was conducted by using the C-AFM mode equipped with TUNA-2 module, capable of measuring junction current in the range of $1\ \mu A$ to $1\ fA$. The reduction was performed by applying a negative bias (typically between $-2\ V$ to $-5\ V$) to the substrate while keeping conducting probe at virtual ground. Change in the local electrical property was measured by recording the current map and local conductance ($dI/dV|_{V_{dc}}$) map (CMAP) simultaneously with the topography under low dc bias (typically $100\ mV$). The $dI/dV|_{V_{dc}}$ maps were recorded by modulating the applied dc bias ($V_{dc}$) with an ac bias ($V_{ac} < 5\%$ of $V_{dc}$) using a signal generator. All CMAPs were recorded with $V_{dc} = 100\ mV$, unless otherwise mentioned, by detecting the in-phase ac output voltage ($\equiv$ ac current signal) using a lock-in amplifier.[35] C-AFM experiments were conducted with Mikromasch Cr-Au coated conducting probes (HQ: CSC37) having force constant $1 - 2\ N/m$ while the tapping mode experiments were performed with NT-MDT NSG-30 probes having resonance frequency in the range of $240 - 440\ kHz$ and stiffness of $40\ N/m$. Nanoscale surface potential mapping was undertaken using an Asylum Research MFP – Infinity AFM in the Kelvin probe force microscopy (KPFM) mode, which allows measurement of surface potential through operation in a two-pass operation. For KPFM measurements, Pt-Ir coated Si tips (Nanosensors PPP-EFM) with stiffness of $2.8\ N/m$ and resonance frequency $\sim 75\ kHz$ were used. Lateral bias was applied in the tapping mode for potential mapping across the reduced GO region. The Raman spectroscopy measurements were conducted using HORIBA XPLORA PLUS micro-Raman set up. Raman mapping was performed with a spatial resolution of $0.7\ \mu m$ using $532\ nm$ excitation ($0.6\ mW$ power) through a 100X objective with $NA = 0.9$ resulting in an illumination spot size $\sim 720\ nm$. Temperature-dependent transport studies were conducted using a closed-cycle cryostat (ARS Inc.) interfaced with a custom written LAB-View program controlling a Keithley 2400 source-meter and Lake Shore 336 temperature controller.

## RESULTS AND DISCUSSION

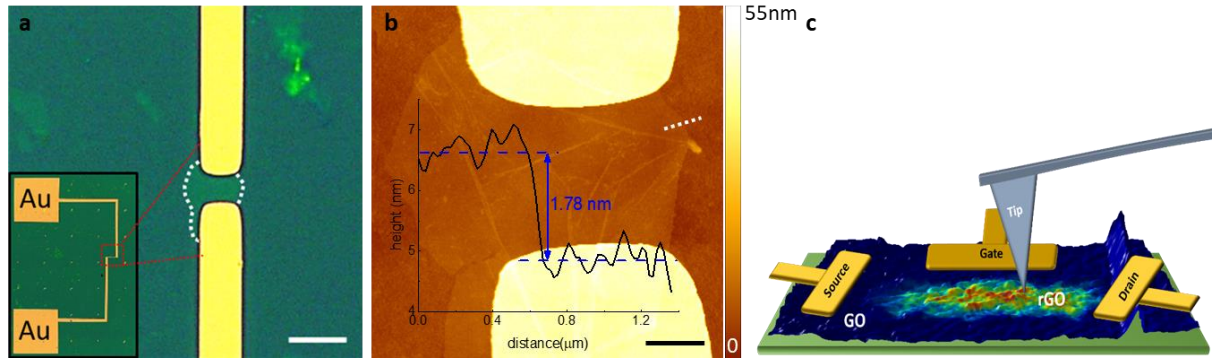

**Figure 1.** Scheme of device fabrication and GO reduction. (a) Optical image of a source-drain terminals on a GO flake (dotted line) with scale bar: 10 $\mu m$. Inset shows low magnification image of Au-pads. (b) AFM topography of the same device overlaid with line scan showing height along the dashed line across the edge of monolayer flake. Scale bar: 2 $\mu m$. (c) Schematic of AFM tip-based reduction technique and proposed device architecture. Colour contrast shows a real change in local conductance induced by the tip-based reduction.

Figure 1a shows an optical image of the GO device electrode layout with the GO layer demarcated by a dashed line. Figure 1b shows AFM topography of the device before reduction of GO along with a line scan across the edge of the GO flake indicating a monolayer with an average thickness of 1.78 nm. Figure 1c shows a schematic of the AFM tip-based nanoscale writing configuration and architecture of a three-terminal device. The surface colour map denotes experimentally recorded variation in local conductance of the GO flake. The initial tip-based reduction experiments were conducted with the GO flakes deposited onto conducting substrates (Au or ITO) with an optimal sample bias. Repeated reduction of GO flakes evidenced stable, uniform and controllable reduction under negative sample bias around $-3\ V$, which was used as a standard through the study. Figure 2a shows a 2 $\mu m$ × 2 $\mu m$ 3D height image of a rGO/GO film overlaid with the current map on the recorded at $-100\ mV$ dc bias, showing the variation of local current 0 $nA$ to $-1.9\ nA$ between the GO and rGO region (500 $nm$ × 500 $nm$ at the centre). The large current carried by the reduced region provides an indication of the change in local electrical property of the rGO region. Evolution of the local electrical properties were further investigated by recording of sequential point $IVs$ across a junction, as a function of voltage cycles obtained by ramping the sample bias between $\pm 3\ V$ as shown in Figure 2b. Figure 2c shows the bias dependence of junction conductance ($G = dI/dV$) obtained by numerical differentiation of selected $IVs$. The zero-bias conductance ($G_0$) gives a good measure of the change in local electrical properties with number of cycles of bias ramps, as plotted in Figure 2d for three different samples evidencing the systematic increase in $G_0$ with repeated bias cycles. Absence of measurable current in the first few bias cycles, and the absence of any non-zero current (and thus $G$) between $\pm 1\ V$ in Figure 2c attests to the insulating nature of the virgin GO flake, even though it rests on a conducting substrate. Figure 2d also indicates that while the rate of increase of $G_0$ (degree of reduction) with bias cycles increases rapidly between the $5^{th}$ and $50^{th}$ cycle, it slows down in subsequent cycles with $G_0$ "saturating" thereafter. Together, the $IV$ characteristics and $GV$ plots demonstrate increase in local conductance by over 4 orders of magnitude between the $5^{th}$ and the $100^{th}$ bias cycles, at a single point. The above behaviour is highly reproducible across multiple GO samples as shown in Figure S1 (Supporting Information).

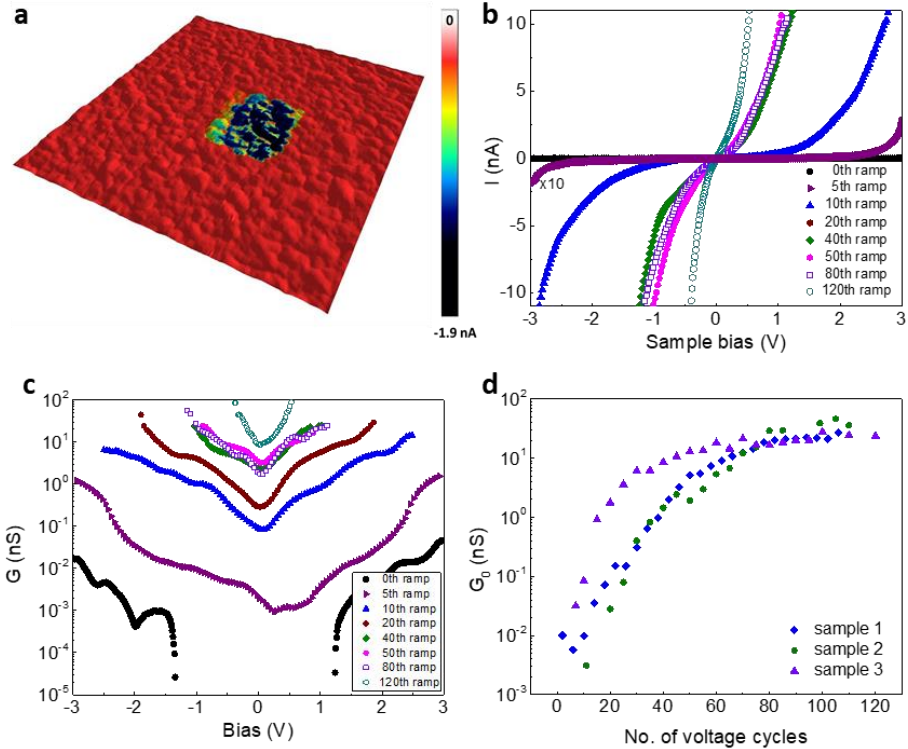

**Figure 2.** Evolution of tip induced GO reduction. (a) 3D height image of GO flake on Au-substrate, overlaid with current map showing the high current carrying reduced GO region at the center. (b) C-AFM junction *IV* characteristics and (c) junction conductance (G) vs sample bias, recorded at different voltage cycle. (d) Variation of zero-bias junction conductance as a function of number of voltage ramp cycle.

Tip induced nanoscale anodic oxidation is a well-known nanolithography technique applied on metals, semiconductors and even graphene.[36, 30-31] Where the key role is played by the presence of a water-bridge formed in the nanoscale tip-sample junction of a SPM, due to capillary condensation[37,31,38] under ambient conditions. It has been suggested, that the high electric field at such biased junctions facilitate the formation of water-bridge and play a dominant role in reduction of GO.[14, 31, 39] Our investigation into the effect of environmental humidity on the tip assisted GO reduction showed that higher relative humidity (RH) favours reduction. Figure S2a and S2b in Supporting Information shows the variation in local current and conductance ($dI/dV$) map across different regions reduced with RH varied between 20% to 80%, on GO. Below RH of 20%, no reproducible change in local electrical properties were observed even for sample bias $-5\,V$ and repeated cycles of reduction. All results presented subsequently have been recorded with RH at 55 ± 5%. These initial results together indicate that the GO reduction process discussed here, as quantified by change in local electrical properties is determined by the following experimental control parameters, sample bias, RH and time, the latter quantified by the number of reduction cycles. Tip induced reduction of GO not only allows spatial confinement of reduction over customizable patterns but also allows a degree of control over tuning the conductance of the rGO region by controlling the above three parameters. The process has significant implications in nano/microscale device fabrication.[30]

Having standardized the reduction protocol in terms of the aforementioned control parameters, GO flakes were immobilized on insulating 300 nm $SiO_2$ on *p*-doped Si substrates and subsequently contacted with electrical pads as shown in Figure 1a,b. Connecting the two electrodes in the source-drain configuration, a rGO channel was patterned by repeatedly scanning the tip over the designated area, as shown in Figure 1c. During the process, the two electrodes were shorted and connected to $-3\,V$ bias used for reduction. Figure 3a shows the conductance map recorded in the device area, clearly demarcating the source-drain electrodes and the interconnecting conducting channel of rGO. The $dI/dV$ map shows that the reduction process induces an increase in average local conductivity by a factor of 10 after 5 reduction scans, which is reproducible keeping the macroscopic experimental parameters fixed. Though some variation in conductivity change has been observed based on the quality of the immobilized GO flake, which are difficult to quantify. Though the change in the electrical properties between the "reduced" and virgin GO regions are quantifiable and directly evident, delineating the chemical identity of the regions remains unclear. The pathway to the defect formation also matters in the electronic conduction adopted by the defect state.[40] If the higher conducting regions are indeed rGO then their chemical identity would necessarily be different than that of GO. Some indication of the change in this chemical identity of the "reduced" region is obtained from the AFM tapping mode phase image shown in Figure 3d, which evidences distinct contrast between the region that had been repeatedly scanned under high negative bias and the "un-scanned" GO region. This gives the first direct indication of change in material property or chemical identity of the "reduced" region vis-à-vis GO.[41-42] GO is electrically insulating in nature due to significant disruption of the $sp^2$ network of graphene, upon incorporation of the various oxidizing groups and consequently, the carbon atoms in a GO layer would show amorphous or disordered arrangement along with interspersed $sp^2$ and $sp^3$ domains,[43-44] as shown in Figure 4a. Absence of long-range order in GO and its insulating nature nominally confirms absence of $sp^2$ domains or if locally present they do not form a percolating network across the GO surface. Generically, reduction of GO to rGO would remove the oxidizing groups and increasingly restore the $sp^2$ network[43, 45] by local nucleation of C=C bonds and restoration of six-fold C rings, as shown in Figure 4b,c. For the tip induced reduction the junction electric field in presence of the water bridge likely leads to local "removal" of the oxygenated functional groups[46] and is accompanied by delocalization of the associated electron cloud, albeit only over the reduced region and increasing local conductance. At this stage, it remains unclear whether "removal" results in complete desorption of the offending functional groups from the surface or possibly their surface migration/diffusion away from the scanned region, under the action of the junction field etc. To estimate this extent of local nucleation of C=C bonds and C rings, micro-Raman spectroscopy and mapping was employed.[47-49] Evolving signatures of both, C=C $sp^2$ hybridized carbon and the C rings offer the first indicators towards restoration of the hexagonal network of graphene (Figure 4d). Figure 3c shows the optical image of a two-terminal monolayer GO device, with the red rectangle indicating the reduced region and the green rectangle ($8 \ \mu m \ \times \ 1 \ \mu m$) between the electrodes denoting our area of interest corresponding to the data presented in Figure 3d. The Raman spectra acquired at two spots each in the "reduced" and pristine GO regions are shown in Figure 3e. All the spectra are normalized with their respective intensity ($I_D$) at the D-band peak (~$1330\ cm^{-1}$). Spectra prior to normalization are shown in the inset. The normalized spectra show that the G-band (~$1580\ cm^{-1}$) intensity ($I_G$) is relatively weaker, compared to the D-band intensity, for spectra recorded on the "reduced region" than on pristine GO. This is more clearly shown in the series of images in Figure 3d which plots the spatially resolved conductance map and $I_D/I_G$ ratio along with fwhm of the G ($\Gamma_G$) and D ($\Gamma_D$) bands, acquired within the green rectangle in Figure 3c. The D and G peaks are well-known Raman signatures corresponding to the in-plane stretching of sp$^2$ C=C bonds and symmetric breathing of standalone six-fold carbon rings, respectively. The latter also quantifies the degree of order/disorder within an agglomeration of C atoms.[50-51]

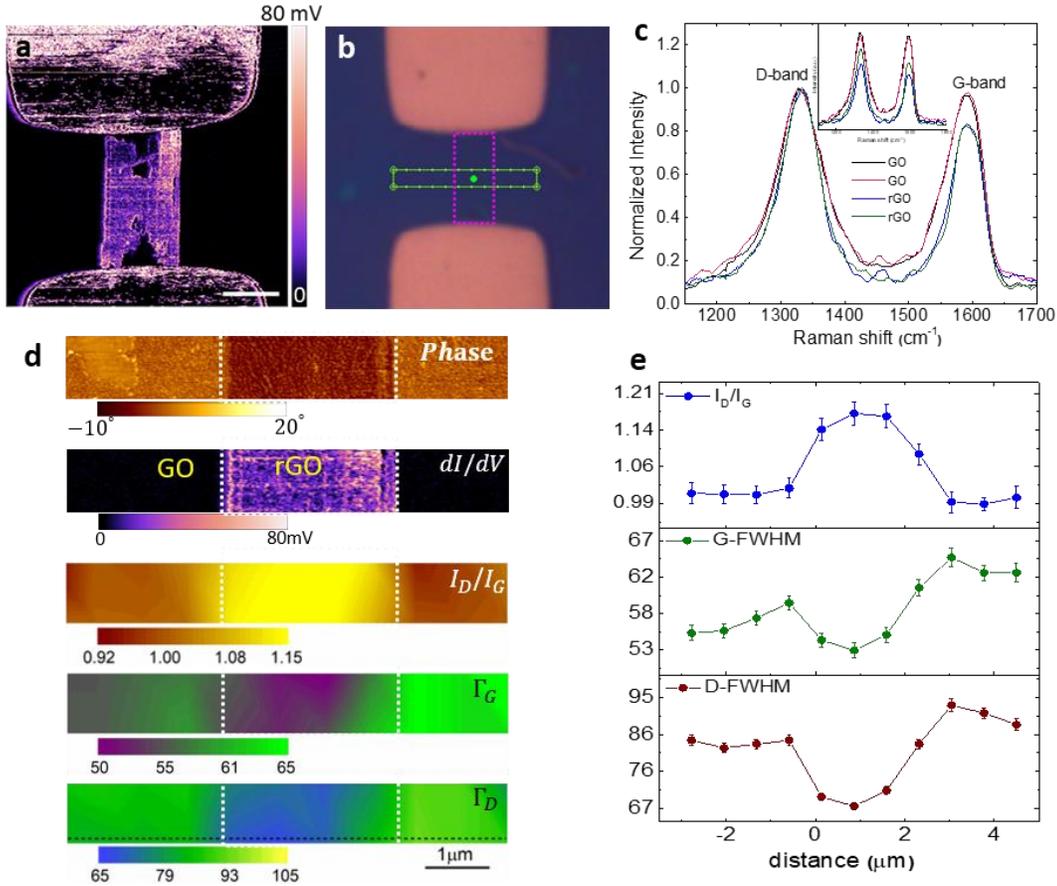

**Figure 3.** Correlating spatially resolved conductance and Raman maps across GO-rGO regions. (a) Conductance map of rGO device. Scale bar: 2 $\mu m$. (b) Optical image of device, green box shows the area of interest and magenta box shows approximate boundary of the reduced region. (c) Normalized Raman spectra at various spots over "reduced" GO and pristine GO regions. Inset shows same spectra prior to normalization. (d) Spatial maps in the area of interest showing phase and conductance ($dI/dV$) contrast, $I_D/I_G$ ratio, $\Gamma_G$ and $\Gamma_D$ between the rGO region in the middle and GO regions on either side. Vertical dashed line indicate boundary of the scanned region. (e) Averaged line-scan across the $I_D/I_G$ ratio, $\Gamma_G$ and $\Gamma_D$ maps in (d).

Assuming the arrangement of C atoms in the native GO to be highly disordered (see discussion on Raman linewidths below), the tip-induced reduction process would increase local order leading to formation of graphitic domains, increase in-plane electrical conductance above a critical fraction, and overall size of the graphitic domains (see Figure 4d). The $dI/dV$ map in Figure 3d shows clear contrast between the pristine GO and the reduced rGO region as do the $I_D/I_G$, $\Gamma_D$ and $\Gamma_G$ maps with a high degree of spatial registry across the four maps. Line scans taken along the length $I_D/I_G$, $\Gamma_D$ and $\Gamma_G$ maps (and averaged across their width) are shown in Figure 3f. While the $I_D/I_G$ value on native GO varies between 0.99 ~1, the tip modified section (rGO) shows an increased value in the range of 1.14 ~1.17. The $I_D/I_G$ ratio has been used earlier to quantify defects and disorder in graphene and other related carbon networks.[47-48, 50] While increase in $I_D/I_G$ ratio in crystalline graphene quantifies increase in defect density (disorder) the opposite may be concluded for dominantly amorphous carbon systems like GO, where increase in $I_D/I_G$ ratio denotes restoration of aromatic six-fold carbon rings[50] signature of reducing defects and increasing order. Here, the increase in the value of $I_D/I_G$ in the reduced region likely indicates increased formation of carbon rings. The increased magnitude of $I_D/I_G$ from 0.99 for GO to 1.17 for rGO, may also be used to estimate a disorder parameter i.e. mean distance between the defects, which increases by ~ 10%.[48] Further confirmation of increased order in the reduced region is obtained from the narrowing of the D and G peak widths shown by the $\Gamma_D$ and $\Gamma_G$ maps (Figure 3d) and line scans (Figure 3e). The large values of $\Gamma_D$ (~ 90 $cm^{-1}$) and $\Gamma_G$ (~ 60 $cm^{-1}$) obtained for the GO samples demonstrates the highly disordered or amorphous nature of the starting material. At the reduced region, $\Gamma_D$ decreases by ~ 20 $cm^{-1}$ and $\Gamma_G$ decreases by ~ 10 $cm^{-1}$. The D and G peak width depends on the phonon lifetime which in turn depends on the size of sp² domains. The decreasing peak width thus also indicates increasing lifetime ($\Gamma \propto 1/\tau$) thus would be indicative of reduced scattering probability from defects.[52] Altogether, the decrease in the parameters $\Gamma_D$ and $\Gamma_G$, increase in the $I_D/I_G$ ratio coupled with the increase in local conductance of the reduced regions shows increase in order i.e. hexagonal sp² carbon rings, and also indicate increase in sp² domain size with decreasing interdomain separation.[53]

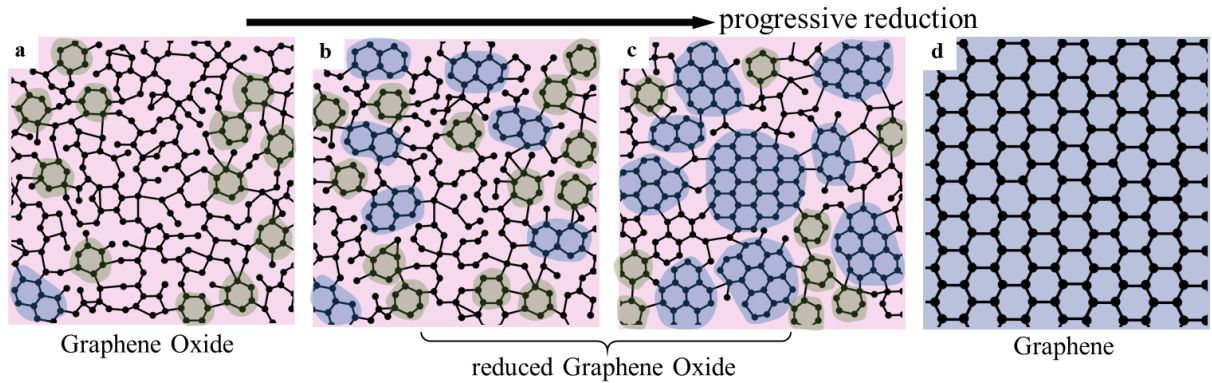

**Figure 4.** Schematic of 2-dimensional arrangement of C atoms on GO as a function of progressive reduction. (a) GO flake showing their amorphous and disordered nature (pink) with few six-fold symmetric C rings (olive), (b) & (c) reduced GO showing increasing local order by the formation of extended graphitic domains (blue) (d) graphene flake showing ideal hexagonal ordering (blue).

The envisaged atomic model of rGO i.e. phase separated regions of graphitic domains, hexagonal C rings etc. embedded within the amorphous carbon background would create a system with highly heterogeneous surface electrical properties. With increasing reduction higher conductive graphitic regions would increase in size and density[40] at the cost of the disordered carbon background, ultimately giving rise to long-range hexagonal order – as depicted in Figure 4. The associated surface inhomogeneity and its role in mediating the lateral electrical transport was probed using scanning KPFM, a surface potentiometry technique, which allows accurate spatial profiling of surface potential. The KPFM measurements rely on the matching of tip bias ($V_{DC}$) with the contact potential difference between the sample and the probe, $V_{CPD}$, by nullifying the vibration of the probe, which is initially driven by the electrostatic force, induced on the AFM probe. We note here that the mapped surface potential in KPFM has contributions arising from differences in work function between tip and sample, as well as the presence of uncompensated surface charges (particularly for non-metallic samples).[54-55] We reiterate that the rGO devices investigated here were fabricated on GO flakes immobilized between two pre-patterned Au probes on 300 nm $SiO_2$ on Si ($p^{++}$). As shown in Figure 5a, the conducting Si substrate and one Au contact were grounded with lateral dc bias ($V_L$) applied to the other Au contact. Figure 5b shows the surface potential ($V_S$) variation across a typical rGO channel, in the absence of any externally applied $V_L$ (= 0), the colour bar showing a potential range of ~ 60 mV. Figure 5c shows the surface potential map of the rGO channel for various $V_L$ varied between 0 – 1 V. All the 3-dimensional plots have been shown with a fixed vertical axis range but with the colour map range varying with $V_L$ for clarity (see Figure caption). Expectedly, all potential maps for non-zero $V_L$ show linear potential variation along the direction of the applied potential (*x*), i.e. length of the rGO channel. Note that at the source and drain electrodes, where the rGO channel is in contact with the underlying Au electrode, $V_S$ shows no variation along *x*-direction. Between the electrodes, the rGO channel assumes a potential commensurate with the applied $V_L$. The difference between the applied $V_L$ and the measured $V_S$ likely originates from a contact resistance within the electrical circuit. The $V_S$ map corresponding to $V_L = 0 \, V$ with $\Delta V_S$ range ~ 60 mV reflects the spatial variation of the tip – surface contact potential difference ($V_{CPD}$). The $V_{CPD}$ for metal-metal junctions quantify the difference between the work functions of the tip and sample, the same interpretation may be extended to a junction between a metal tip (Au) and a gapped semiconductor or semi-metal sample (rGO). Here the sample work function ($W_S$) is given by $W_S = E_C + \chi - E_F - eV_B$,[56] where $E_C$ is the conduction band minimum, $E_F$ is the Fermi level, $\chi$ is the electron affinity and $eV_B$ quantifies any band-bending due to surface effects, such as charges, unsaturated bonds etc.

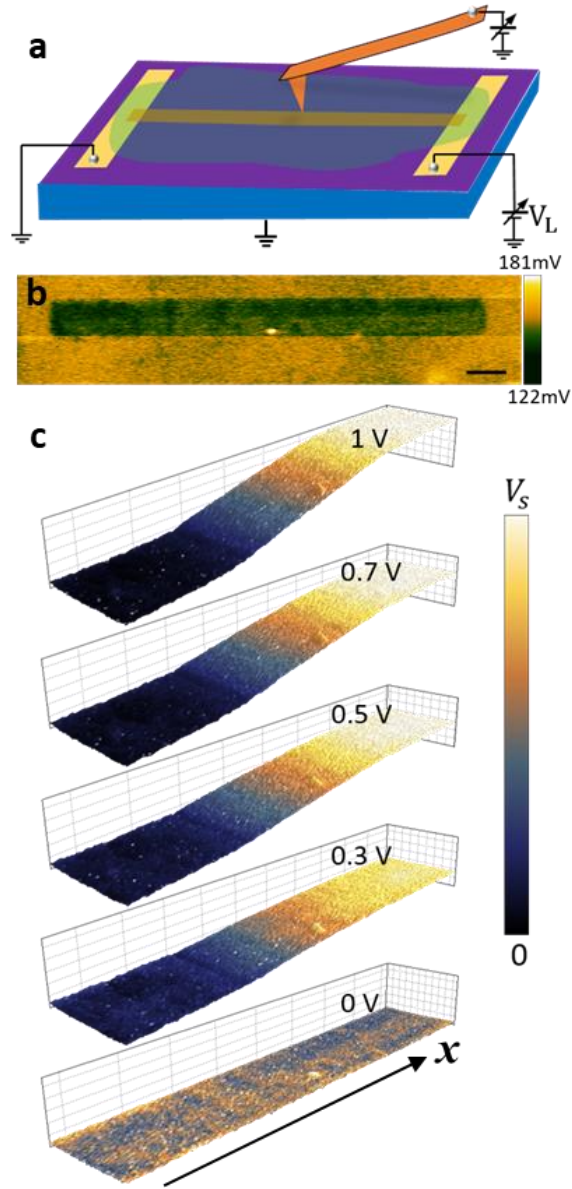

**Figure 5.** Change in local surface potential due to reduction in the presence and absence of external dc electric field. (a) Schematic of Scanning-potentiometry. (b) Surface potential of a rGO-channel without any bias. Scale bar: 2 $\mu m$ (c) Variation of channel potential as a function of different bias ($V_L$) applied to the right electrode while keeping the left electrode grounded. Colour bar represents the variation of surface potential in the range 0 to $V_s$, Where $V_s$: 572 $mV$ ($V_L = 1\ V$), 400 $mV$ ($V_L = 0.7\ V$), 300 $mV$ ($V_L = 0.5\ V$), 200 $mV$ ($V_L = 0.3\ V$), 60 $mV$ ($V_L = 0\ V$).

Thus the zero bias potential map in Figure 5b indicates that the GO region has a higher $V_{CPD}$ with respect to the tip (Pt) $E_F$, which is held at 0 $V$ (virtual ground), compared to the rGO region. The $V_{CPD}$ difference arising from a difference in the $\chi$ and $eV_B$ between GO and rGO regions. This experimental finding is qualitatively commensurate with the equilibrium band diagrams at the GO−rGO interface, where GO has a larger bandgap compared to rGO, but this does not allow us to extract values for the individual band gaps. In interpreting the contrast in the potential maps (Figure 5c), we assume that these rGO channels form a composite 2D heterostructure. Where the disordered GO background, modelled as a bandgap semiconductor, embedded with the more conducting and ordered graphitic domains, akin to the schematic shown in Figure 4b,c constitute the rGO[24, 34]. Thus the rGO surface is a composite, composed of materials with spatially varying band properties and doping, nucleating a 2D array of nanoscale junctions at the domain boundaries. Indeed it can be shown that the spatial contrast of the potential map would be theoretically given by $\delta V_S(x,y) = \delta E_C(x,y) + \delta\chi(x,y) - e\delta V_B(x,y)$.[56] For the $V_L = 0\,V$ case, thermodynamic equilibrium across the rGO surface i.e. a spatially invariant $E_F$, is established by charge transfer and band-bending across the domain boundaries giving rise to the local potential variation, strong boundary electric fields and localized charges. Figure 6a shows the quiver plot of the local electric field, calculated for a section of the potential map for $V_L = 0\,V$. The direction and strength of the electric field is represented by the orientation and length of the arrows. The stronger fields localized at the domain boundaries demarcate the dominant scatters that would impede electrical transport along the rGO channel. Figure 6b shows the potential map and quiver plot for $V_L = +1\,V$ (applied to the right electrode keeping the left electrode at ground potential), showing a dominantly linear potential drop along the length of the channel with the local fields primarily pointing along the –ve x-axis. Figure 6c again plots the surface potential for $V_L = +1\,V$, but after subtraction of the local dc potential i.e. linear plane corrected, and the corresponding local electric field.

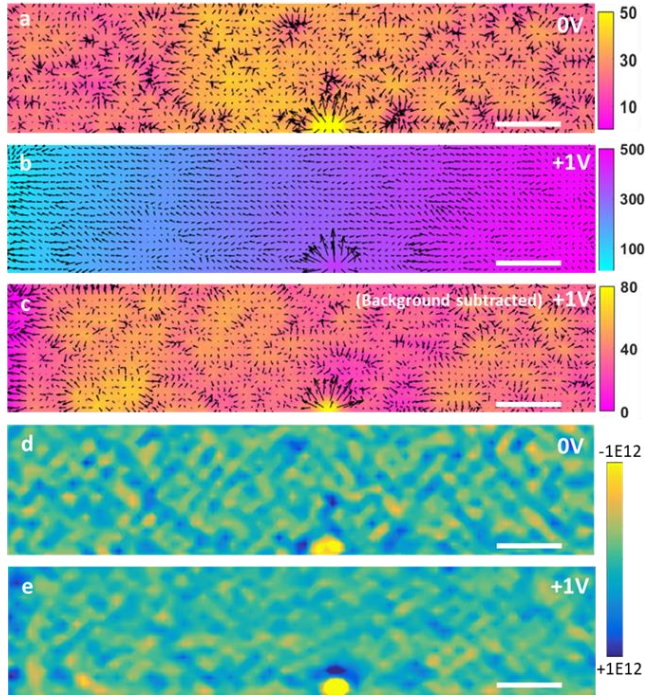

**Figure 6.** Spatial inhomogeneity of surface potential, electric field and charge across a rGO surface. Spatial maps of local surface potential (VS) for a section of rGO channel, arrows indicate local electric field for (a) $0\,V$ applied across the channel; (b) $+1\,V$ applied to the right-electrode, left-electrode is at zero; (c) same as (b) after subtraction of the linear potential gradient. Direction and length of arrows represent the direction and relative strength of local electric field. colour-bar shows variation of $V_S$ in mV. (d) & (e) shows mapping of $\nabla^2 V_s$ for $0\,V$ and $+1\,V$ applied external bias between two electrodes respectively. Unit of $\nabla^2 V_s$ represented by Colour-bar is V/m², Scale-bar: 1 μm.

The sample corresponding to the data shown in Figure 5 and 6 had a rGO channel length of $11.5\,\mu m$ and width $2.5\,\mu m$. For the $V_L = 1\,V$ the measured $\Delta V_{CPD}$ between the source-drain electrodes was $\sim 550\,mV$ which corresponds to macroscopic dc electric field of $4.1 \times 10^4\,V/m$ along the channel. By comparison, the local fields obtained from the experimental potential maps have a wide distribution in magnitude, varying between $10^2$ V/m to $10^6$ V/m, with a mean value of $4.3 \times 10^4\,V/m$, close to the calculated macroscopic field. The large difference between the macroscopic and microscopic field values likely plays a dominant role in determining the nature of electrical transport in these systems and explain the observed variation in VRH transport presented in earlier reports[24, 27]. Higher spatial inhomogeneities in the potential map would correspond to both higher local fields and spatial non-uniformity with respect to the macroscopic mean value. The measured surface potential $V_S(x, y)$ would satisfy the Poisson equation given by $\vec{\nabla}.[\epsilon(\vec{r})\vec{\nabla}V_S(\vec{r})] = -\rho(\vec{r})$ where $\epsilon(\vec{r})$ is the spatially dependent permittivity and $\rho(\vec{r})$ is the local charge density. A few-layer rGO sample would have a highly anisotropic and inhomogeneous permittivity, which is rather difficult to properly account for in the mathematical model to calculate $\rho(\vec{r})$. However, even neglecting the spatial dependence of $\epsilon(\vec{r})$ the 2D surface plot of the quantity $\nabla^2 V_S$ provides useful information regarding localization of charges on the rGO surface and their associated length scales. Figures 6d,e shows the $\nabla^2 V_S$ plots for the case of $V_L = 0\,V$ and $1\,V$, corresponding to those shown in Figure 6a,b. The images for the $V_L = 0\,V$ case show that $\nabla^2 V_S$ changes sign (signifying local positive and negative charges) over length scales of 80 – 100 nm, which is similar to that seen for the case of $V_L = 1\,V$. Significantly, the *rms* surface roughness of the plots decreases from $2.96 \times 10^{11}\,V/m^2$ to $2.12 \times 10^{11}\,V/m^2$ between $0\,V$ and $1\,V$ bias. Thus the drift-diffusion current resulting from application of the external bias decreases the surface roughness – in effect "smoothening" the charged surface and thus reducing scattering and decreasing channel resistance at higher bias (see discussion on $IV$ data below). As mentioned earlier the spectral features of D and G peaks obtained in the Raman spectra of rGO and graphene provides information on spatial defect distribution[47] and graphitic domain size in the system. Previous investigations on chemically reduced rGO have reported typical $I_D/I_G$ ratios in the range 1 – 2 corresponding to graphitic domain sizes less than 10 nm,[7, 57] which is also corroborated in atomically resolved topographic data[34] and electrical transport measurements[23]. In the present case for rGO with $I_D/I_G \sim 1.17$ (Figure 3c) would likely yield a similar estimate for the localization length scale, which is contrary to the fluctuation length scales seen in the $\nabla^2 V_S$ plots above. However, such "charge puddles", with lateral extent over tens of nanometers have been reported in graphene,[5, 58-60] rGO[5] and Dirac semimetals[61-62] in which charge impurities arising from surface adsorbates, substrate and defects lead to large spatial charge fluctuations, especially with $E_F$ close to the Dirac point. The spatial extent of the observed charge puddles are primarily decided by two factors, background permittivity and the free carrier density which together determine the lateral screening length ($\xi$). The typical carrier density in rGO may vary between $10^9 - 10^{12}$ $/cm^2$ [32, 63] with a wide variation in relative permittivity observed between 3 – 1,000,[64-66] even influenced by environmental conditions. Both these parameters are also spatially variable on the inhomogeneous rGO surface and strongly influenced by the substrate and the environment.[29, 58, 60] Thus even though the disordered domains (harbouring the charged impurities) may be localized over a few nanometers the resulting charge localization or the screening distance may extend over significantly larger length scales, as seen here.

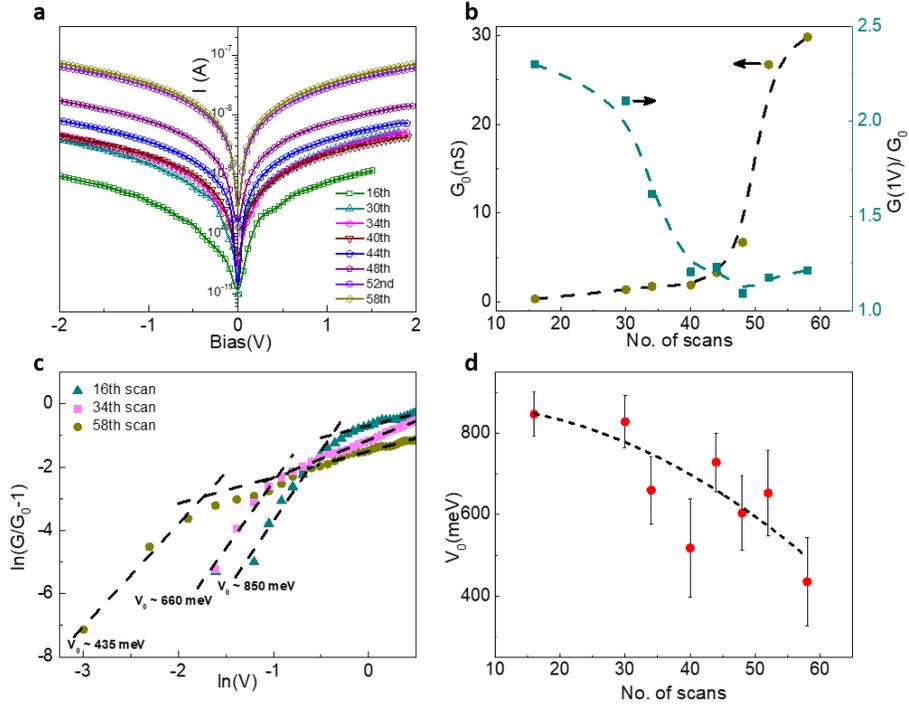

**Figure 7.** Non-linear charge transport and its evolution with progressive reduction of rGO. (a) Lateral *IV* characteristics of a two-terminal rGO devices at different states of tip induced reduction. (b) Zero bias conductance as a function of no. of scans presented in the left axis, the right axis shows the device nonlinearity represented by $G(1\,V)/G_0$ as a function of no. of scans. (c) $\ln(G/G_0 - 1)$ vs. $\ln(V)$ plot at different reduction state of the rGO device. Parameter $V_0$ was calculated from the linear fit in the low bias regime of the plot c. (See in the writing section) (d) Variation of $V_0$ with no. of scans performed for reduction of GO.

As mentioned earlier repeated scanning of the rGO channel by the C-AFM tip (at fixed -3 V sample bias) increases the local conductance within the channel and also increases the overall lateral conductance of the channel, finally saturating to a steady-state value. Typically, it was possible to increase the lateral channel conductance by $\sim 10^4$ times by repeatedly scanning the same area by more than 60 times reaching a steady-state value. Experimentally, larger number of reduction scans result in physical damage to the rGO layers and hence avoided. Figure 7a shows the semi-log plot of lateral $IV$ characteristics obtained for the device, where the designated area between the source-drain contacts was repeatedly scanned with the C-AFM tip under -3 V tip-sample bias for reduction of GO. The non-linear $IVs$ are representative of electrical transport through highly disordered systems, where tunnelling and/or hopping between localized states or spatially segregated electron-rich domains (Figure 4b,c) gives rise to channel conductance. Electrical transport in rGO has been previously reported to be described by various 2D-VRH models[24, 63] as well as by thermally activated transport[22, 63], similar to experimental observations in graphene.[67,6, 27] There is limited understanding about the exact physical parameters that bring about each transport mechanisms, their domain of applicability and importantly the consequences of such identification of transport model. Figure 7a also indicates that the channel conductance ($G$) systematically evolves with progressive reduction cycles – quantified by the number of tip scans performed. Figure 7b shows that the zero-bias conductance, $G_0$ increased by three orders in magnitude between the 16$^{th}$ – 70$^{th}$ scan, which is the typical evolution observed across the various devices investigated. Along with the increase in $G_0$, the bias dependent nonlinearity decreases significantly with increasing number of scans. The normalized dynamic conductance at $1\,V$, i.e.

$G(1\,V)/G_0$ (Figure 7b right axis plot) decreases from 2.3 (16th scan) to 1.2 for the 58th scan and indicates progression towards linear or increasingly ohmic transport along the channel. To analyse the evolution of electrical transport along the channel with reduction we obtained the conductance spectra, $G\ (=dI/dV)\ vs.V$ by numerically differentiating the $IV$ characteristics, which were fitted to an empirical formula given by,

$$G(V) = G_0 \left(1 + \left|\frac{V}{V_0}\right|^n\right) \quad (1)$$

where $V_0$ and $n$ are fit parameters. Figure 7c plots $\ln(G/G_0 - 1)$ vs. $\ln(V)$, for the $IV$s recorded after 16th, 34th and 58th scans, where the slope and intercept yield the parameters $V_0$ and $n$ of the equation above. Evidently, a single exponent does not fit the entire bias range and in all cases show a distinct cross over, demarcating high and low bias regimes. The linear fits in these regimes show that $n \sim 1\ (\pm 0.2)$ for the high bias regime and $n \sim 4\ (\pm 0.5)$ at lower biases, across all $IV$s studied. The bias dependence also indicates that close to zero bias the conductance has a significantly stronger bias dependence than in the high bias regime. In all cases, the final device has the lowest resistance. However, even the most conducting device still exhibits non-linear conduction resulting in significantly lower device resistance at higher bias ($\geq 1V$). This is consistent with the decrease in rms roughness of the $\nabla^2 V$ surface plots between 0 V and 1 V seen earlier. It is likely that a part of the observed nonlinearity arose from non-ohmic nature of transport at the terminal contacts, which is the residual non-linearity retained in the most reduced devices. Experimentally, the $IV$s recorded after the first few tip scans, close to the noise floor of the measurement system, showed voltage offsets akin to charging between the source-drain electrodes of the order of few 10s of mV. The offsets systematically decreased after increased number of scans, becoming non-existent after the 30th scan.

The above observations are again suggestive of a disordered conductor with a depleted DOS at its $E_F$, which progressively fills up with increased number of scans. That is, for pristine GO its $E_F$ lies within its bandgap ($E_G$) with zero DOS at $E_F$, and with progressive reduction scans result in addition of localized states around its $E_F$, increasing its graphene-like character. For GO the typical energy required for the carriers to become delocalized and conduct (i.e. the activation energy) is determined by $E_G$, but the band diagram gets complicated for the inhomogeneous rGO. For rGO, carriers need to overcome a localization energy quantified by the parameter $V_0$ in equation 1. Figure 7d plots the decrease in the best fit values of $V_0$ with increased number of scans, indicating that the localization energy the carriers have to overcome decreases progressively. As mentioned earlier, previous investigations on temperature dependence of conductivity identify electrical transport in rGO to be driven by Mott-VRH, ES-VRH and activated transport, however, the domain of validity of such identification and more importantly the consequences of such identification remain unclear. Our results show that devices with lower final values of rGO channel conductance (more insulating, $R \sim 300\ k\Omega$) show thermally activated transport while those with higher final conductance (less insulating, $R \sim 50\ k\Omega$) follow the Mott-VRH model for 2D systems for the temperature range investigated. Figure S4 evidences linear dependence of $\ln G_s$ on 1/T, for the low conducting devices (D1 and D2), indicating Arrhenius type fits yielding activation energies $E_A$ values $166\ meV$ and $138\ meV$. In contrast, the more conducting devices (D3 and D4) show a linear dependence of $\ln G_s$ on $1/T^{1/3}$. See Supporting Information for further details. The limited temperature range explored for the more insulating samples were borne out of the limitations in the maximum measurable resistance in the experimental setup. The transport data presented above indicate that in rGO conduction would be mediated by carriers hopping across localized states near the $E_F$, lying within the bandgap of the pristine GO. The density of such localized states, their energies ($\epsilon_i$) and occupation being determined by the degree of reduction of GO i.e. the number of scans. Electrons in these localized states have a hopping probability determined by the spread in $\Delta\epsilon_i$ vis-à-vis spatial separation between the states (domain size ~ $r$),[25] which are all dynamically controlled by the degree of reduction. The *IV* analysis above yields an effective localization potential in $V_0$, which is the cumulative response across the highly inhomogeneous rGO system, harbouring the graphitic domains in the GO background. Our present results lack sufficient evidence to draw more detailed inferences but to conclude that the rGO system would demonstrate VRH like transport or activated transport dependent on the bandgap, the temperature and the energetics of the localized states close to the $E_F$. The large distribution in the local electric field values also indicates that more than one type of VRH may be contributing to the local transport.

CONCLUSIONS

In conclusion, we have investigated fabrication of reduced GO devices, individually written onto insulating GO flakes by nanoscale local reduction using a C-AFM tip. The reduction process offers a high degree of control and standardization through three macroscopic parameters, junction bias, relative humidity and number of reduction cycles, controlling the degree of reduction. Change to the morphological and electronic properties of the GO surface after reduction, investigated using spatially resolved electrical measurements and micro Raman spectral mapping provided conclusive evidence towards reduction of GO in the designated area. Electrical transport through the two-terminal rGO channels is non-linear, which correlates well with the spatially resolved two-dimensional surface potential maps. The variation reflects the inhomogeneous electronic structure across the surface, and segregation into local positively and negatively charged regions. The current-voltage characteristics showed systematic decrease in non-linearity with progressive reduction of the channel, indicating that the nonlinearity originates from the material properties of rGO, rather than from transport across the terminal metal-rGO contacts. Together, temperature and bias dependent transport, and the conductance and potential maps give further evidence of charge localization effects close to the Fermi level with multiple transport mechanisms contributing to the overall conduction. Depending on the degree of reduction and the disorder in the system, electrical transport in rGO may be dominated by thermally activated or Mott-VRH, or carry signatures of electric field driven variable range hopping, due to the high local fields that are significantly stronger than the applied average field. The findings not only provide for better understanding of the nanoscale electronic structure of rGO and electrical transport but also demonstrates the scope of tuning its electronic and electrical properties by restoration of the *sp*$^2$ hybridised carbon lattice. It is anticipated that these results will further the scope towards realisation of all graphene electronic devices such as transistors, switches and tuneable infrared detectors leveraging not only the technical control allowed by this scheme but also the functional aspects of the investigation.

---------------------------------------------------------------------------------------------------------------------- --------------

## ASSOCIATED CONTENT

**Supporting Information**.

SI-1, Evolution of junction $IVs$ as a function of voltage cycles; SI-2, Effect of humidity on GO reduction; SI-3, Effect of applied bias magnitude on GO reduction; SI-4, Conductance vs Temperature. (PDF)

## AUTHOR INFORMATION

**Corresponding Author**

# Email: arijit17@iisertvm.ac.in,

† Email: j.mitra@iisertvm.ac.in

**Present Addresses**

* Department of Functional Materials, Łukasiewicz Research Network- Institute of Electronic Materials Technology, Wolczynska 133, Warsaw, Poland.

**Authors contributions:**

J.M., S.R.P.S & K.B. conceived the project. Arijit, H.G., K.B., A.K. and J.M. performed the experiments. Arijit and J.M. have analysed and interpreted the data. All authors have contributed to writing the manuscript.

**Notes**

The authors declare no competing financial interests.

ACKNOWLEDGMENT

The authors acknowledge financial support from SERB, Govt. of India (SR/52/CMP-0139/2012, CRG/2019/004965), UGC-UKIERI 184-16/2017(IC) and the Royal Academy of Engineering, Newton Bhabha Fund, UK (IAPPI_77). A.K. acknowledges financial support from the Engineering and Physical Sciences Research Council (Contract No. EP/S037179/1 and EP/N018389/1). Arijit acknowledges research fellowship from IISER Thiruvananthapuram. H.G. acknowledges DST for INSPIRE fellowship.

# Supporting Information:

**SI-1: Evolution of junction *IVs* as a function of voltage cycles:**

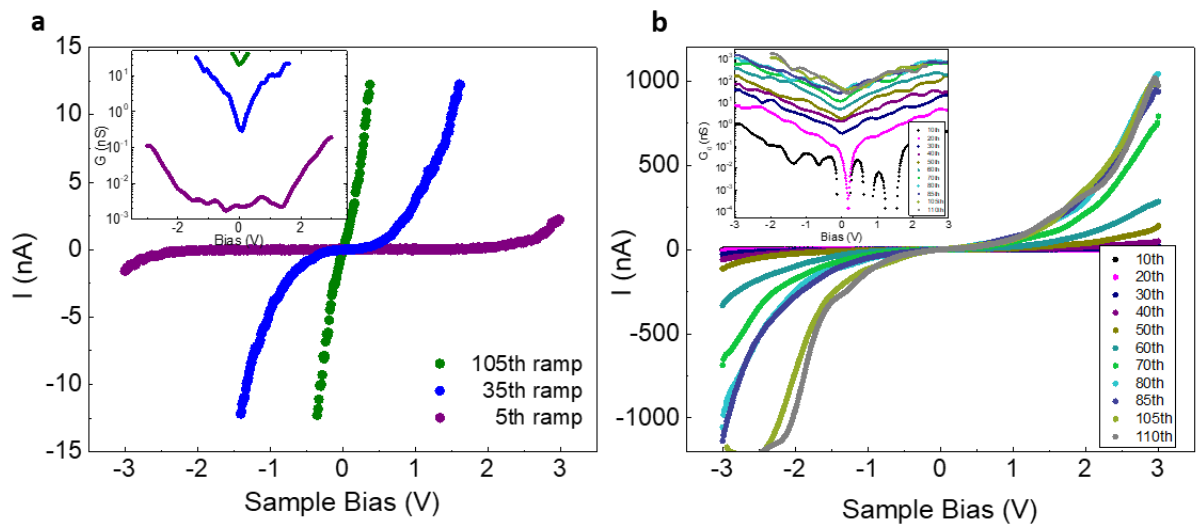

Figure S1. Evolution of tip-sample junction characteristics (a) & (b) Selected point *IV* characteristics of tip (Au) – sample (GO) junctions after repeated voltage ramp cycles of ±3 V, for two different GO samples. Inset: Junction conductance vs. bias of corresponding to the *IV*s. All bias cycles were performed between ±3 V, however, the highest detectable current value was decided by the amplification factor of the current amplifier which was 1 nA/V for (a) and 100 nA/V for (b).

## SI-2: Effect of humidity on GO reduction:

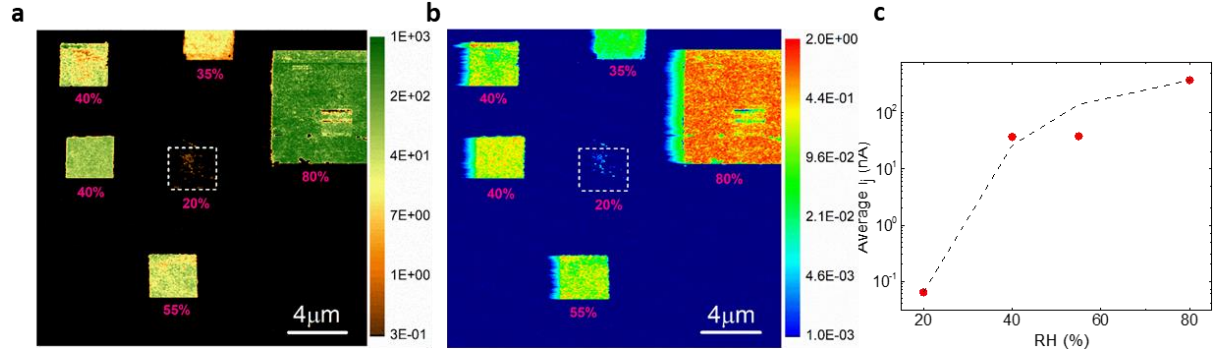

Figure S2. Reduction as a function of relative humidity. (a) Current map (colour map unit : nA) and (b) conductance map (colour map unit: V) for GO (rGO) surface, reduced with – 4 V under different relative humidity (RH%) conditions. (c) Variation of saturated junction current ($I_j$) vs RH% (Ij is averaged over the corresponding reduced area).

The effect of environmental humidity on the reduction of GO was studied by performing reductions with the bias of the same magnitude at different levels of relative humidity (RH). Figure S2a shows the current-map of a GO film acquired simultaneously with topography at +1 V (dc). The squared portions shown in both the current map and $dI/dV$ map were reduced at -4 V at different RH. Whereas the portion reduced at 80% of RH level shows the highest current, the portion reduced at 20% RH shows the almost negligible current. $dI/dV$ map (Figure S2b) acquired simultaneously with the topography, and current-map also shows a similar behaviour of negligible reduction with lower RH. Even the average junction current ($I_j$) measured as a function of RH (Figure S2c) shows increase by a factor of 103 with RH increasing from 20% to 80%, evidencing the critical role played by environmental humidity in the reduction process.

**SI-3: Effect of applied bias magnitude on GO reduction:**

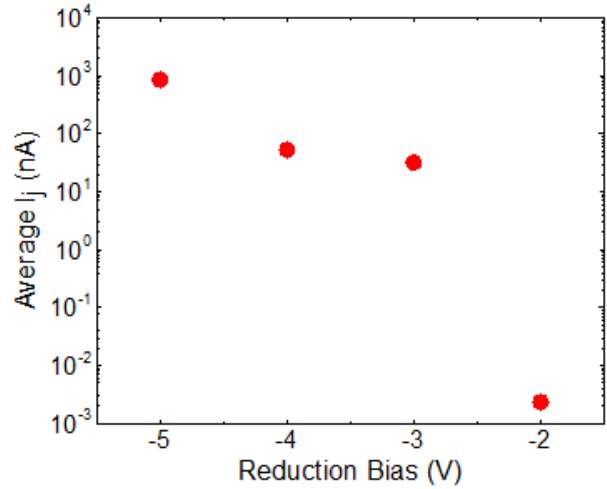

Figure S3. Reduction as a function of applied bias. Average junction current ($I_j$) measured after reduction of GO with different reduction bias by scanning a designated area for single time while keeping the scan parameters identical.

The efficiency of the tip induced reduction process was also studied as a function of the magnitude of applied reduction bias. Electrical conductance increases for the region reduced with higher bias has been observed. Figure S3 shows average $I_j$ as a function of reduction bias magnitude, which indicates average $I_j$ of a reduced area can be increased by a factor of $\sim 10^6$ by suitably varying the reduction bias.

## SI-4: Conductance vs Temperature:

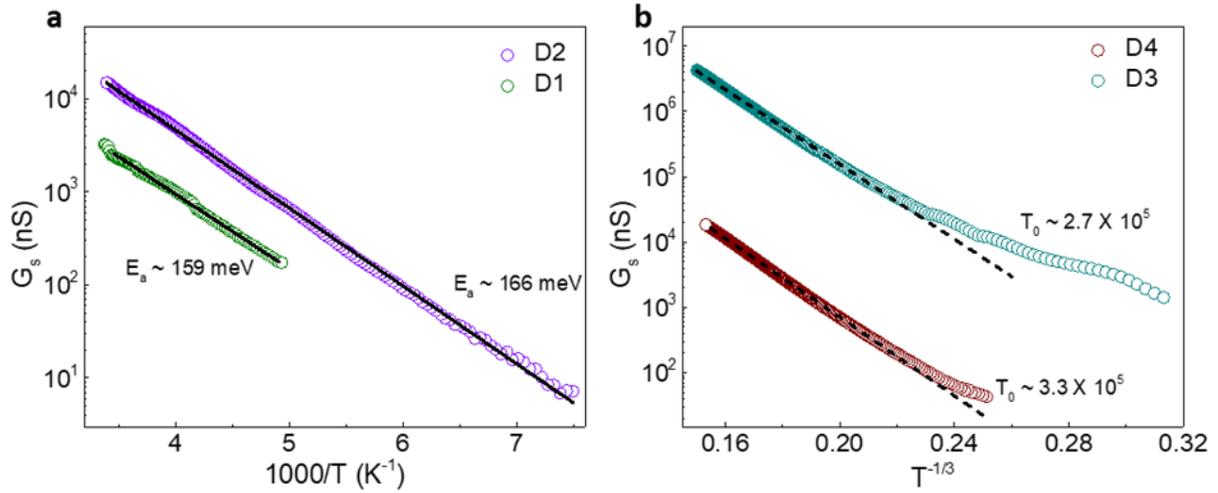

Figure S4. Temperature-dependent transport measurements. (a) Quasi four-probe sheet conductance ($G_s$) $vs.\ 1/T$ for highly resistive rGO devices ($R \sim 300\ k\Omega$). Black solid line shows a linear fit corresponding to thermally activated transport. (b) Sheet conductance ($G_s$) $vs.\ T^{-1/3}$ for lower resistive ($R \sim 50\ k\Omega$) rGO devices. Black solid line shows a linear fit corresponding to 2D Mott-VRH transport.

Temperature dependence of device resistance were recorded to comprehend the mechanism of charge transport through the disrupted sp2 network of rGO. The temperature-dependent data were analyzed using both thermally activated transport and Mott 2D-VRH models. Figure S4a shows sheet conductance (Gs) vs. 1/T plot in logarithmic scale for two lower conductance (R ~ 300 kΩ) multilayer GO-rGO devices patterned with a C-AFM tip. For lower conductance sample $G_s\ vs.\ 1/T$ plots fits well with the thermally activated transport mechanism $\{G_s(T) = G_0 exp(-E_a/k_B T)\}$ and yields the value of activation energy ($E_a \sim 160\ meV$) and indicates a bandlike transport. Whereas samples with relatively higher conductance (R ~ 50 kΩ) (Figure S4b) shows an agreement with 2D-VRH model: $G_s(T) = G_0\ exp\left\{-\left(\frac{T_0}{T}\right)^{1/3}\right\}$ where, $G_0$ is a pre-exponential factor and $T_0$ is the hopping parameter.

GO due to its highly disruptive sp2 network doesn't provide a sufficient number of localized states, which results in a negligible overlap among the localized states and leads to a bandgap opening hence prohibits the carrier delocalization. Reduction of GO partially improves the network by increasing the no of sp2 domains and there sizes which results in reduction of bandgaps and thus allow carrier transport by thermal excitation. For samples with higher degree of reduction a large no. of sp2 domains will be generated within the defective GO network results in a large no. of localized states near the fermi level ($E_F$). This allows charge carriers to find an energetically favourable site to hop from one localized state to another even though it's located farther away. Lowering the temperature allows the carriers to hop only between the states having similar energy which in turn reduces the hopping probability and hence an increase in resistance. Deviation from 2D VRH in the low-temperature regime is most likely due to dominant field-driven hopping.[1]